\documentclass[aps,prd,eqsecnum,preprint,tightenlines,nofootinbib,showpacs]{revtex4-1}
\usepackage{graphicx,amsmath,latexsym}

\newcommand{\bea}{\begin{eqnarray}}
\newcommand{\eea}{\end{eqnarray}}
\newcommand{\be}{\begin{equation}}
\newcommand{\ee}{\end{equation}}
\newcommand{\ub}[1]{\underline{#1}}
\newcommand{\ob}[1]{\overline{#1}}
\newcommand{\Pminus}{{\cal P}^-}

\newcommand{\Pfree}{{\cal P}_{\rm free}^-}
\newcommand{\Pint}{{\cal P}_{\rm int}^-}
\newcommand{\pp}{p^{\prime +}}

\begin{document}

\title{Scalar theories and symmetry breaking \\
in the light-front coupled-cluster method%
\footnote{Based on a talk contributed to the
Lightcone 2013 workshop, Skiathos, Greece, 
May 20-24, 2013.}
}

\author{S.S. Chabysheva}
\affiliation{Department of Physics \\
University of Minnesota-Duluth \\
Duluth, Minnesota 55812}

\date{\today}

\begin{abstract}
We extend the light-front coupled-cluster (LFCC) method
to include zero modes explicitly, in order to be able
to compute vacuum structure in theories with symmetry
breaking.  Applications to $\phi^3$ and $\phi^4$ theories
are discussed as illustrations and compared with variational
coherent-state analyses.
\end{abstract}

\maketitle

\section{Introduction} \label{sec:intro}

The light-front coupled-cluster (LFCC) method was invented~\cite{LFCC}
as a way of solving light-front Hamiltonian eigenvalue problems
in a Fock basis without truncation of that basis.  This avoids 
various complications of Fock-space truncation, such as uncanceled
divergences, spectator dependence, and Fock-sector dependence.
Instead of Fock-space truncation, the LFCC method restricts the
freedom of wave functions for higher Fock states, making them
dependent on the wave functions for lower Fock states through
vertex-like functions determined by nonlinear integral equations.
However, the original development of the LFCC method did not
explicitly include the zero momentum modes needed for 
representation of a nontrivial light-front vacuum.  Here we
summarize recent work that rectifies this deficiency~\cite{LFCCzeromodes}.

Calculations in discrete light-cone quantization
(DLCQ)~\cite{PauliBrodsky}
have shown that spontaneous symmetry breaking in
$\phi^4$ theory can be detected without zero modes
by studying ground-state degeneracy in the massive
sector~\cite{RozowskyThorn,Varyetal}.  An improved
convergence can be obtained if zero modes are
included~\cite{ZeroModes}, but they are not necessary.
In $\phi^3$ theory, where the spectrum is unbounded
from below~\cite{Baym}, DLCQ calculations without
zero modes~\cite{Swenson} require careful extrapolation;
however, inclusion of zero modes makes the unboundedness
obvious~\cite{ZeroModes}.  Zero modes are also
necessary for construction of a physical light-front vacuum 
with which one might compute vacuum expectation
values and associated critical exponents, as well as
the field shift needed in the Higgs mechanism.

As light-front coordinates~\cite{Dirac,DLCQreviews}, we use 
the time $x^+=t+z$ and space $\ub{x}=(x^-,\vec{x}_\perp)$, with
$x^-\equiv t-z$ and $\vec{x}_\perp=(x,y)$.  The light-front energy
is $p^-=E-p_z$ and momentum $\ub{p}=(p^+,\vec{p}_\perp)$, with
$p^+\equiv E+p_z$ and $\vec{p}_\perp=(p_x,p_y)$.  The mass-shell
condition $p^2=m^2$ becomes $p^-=\frac{m^2+p_\perp^2}{p^+}$.

The LFCC method solves the light-front eigenvalue problem
\be
\Pminus|\psi\rangle=\frac{M^2+P_\perp^2}{P^+}|\psi\rangle
\ee
in terms of a Fock-state expansion for the eigenstate, but
without truncation.  We build the eigenstate as $|\psi\rangle=\sqrt{Z}e^T|\phi\rangle$ 
from a valence state $|\phi\rangle$ and an operator $T$ that increases
particle number.  $Z$ maintains the normalization.
The eigenvalue problem can then be rewritten as
\be
\ob{\Pminus}|\phi\rangle=e^{-T}\frac{M^2+P_\perp^2}{P^+}e^T|\phi\rangle
=\frac{M^2+P_\perp^2}{P^+}|\phi\rangle,
\ee
with $\ob{\Pminus}\equiv e^{-T}\Pminus e^T$ an effective Hamiltonian.
We project the eigenvalue problem onto the valence and orthogonal sectors
\be
P_v\ob{\Pminus}|\phi\rangle=\frac{M^2+P_\perp^2}{P^+}|\phi\rangle, \;\;
(1-P_v)\ob{\Pminus}|\phi\rangle=0,
\ee
with $P_v$ the projection onto the valence sector.  The effective
Hamiltonian is constructed from its Baker--Hausdorff expansion
\be
\ob{\Pminus}=\Pminus+[\Pminus,T]+\frac12 [[\Pminus,T],T]+\ldots
\ee

The approximation made is to truncate $T$ and $1-P_v$.  The latter
truncation is to provide no more equations than are needed to
solve for the terms kept in $T$.  The Baker--Hausdorff expansion
can then be truncated without additional approximation.
The truncation of $T$ and $1-P_v$ can be
systematically relaxed by adding more terms to $T$, ordered
according to the number of particles created.  

Without zero modes, every term in $T$ must include at least
one annihilation operator, so that longitudinal momentum is
conserved.  With zero modes, there can be terms in $T$
that involve only zero-mode creation operators, and it is
these terms that can build a physical vacuum from the Fock vacuum $|0\rangle$.
The vacuum sector of a theory is then investigated by
using the Fock vacuum as the valence state and $\sqrt{Z}e^T|0\rangle$
as the physical vacuum.  Here we explore this approach in the
context of $\phi^3$ and $\phi^4$ theories in two dimensions.
Additional discussion can be found in Ref.~\cite{LFCCzeromodes}.

\section{$\phi^3$ theory:} \label{sec:phi3}

The Lagrangian for $\phi^3$ theory is
\be
{\cal L}=\frac12(\partial_\mu\phi)^2-\frac12\mu^2\phi^2-\frac{\lambda}{3!}\phi^3.
\ee
The mode expansion for the field at zero light-front time is
\be \label{eq:mode}
\phi=\int \frac{dp^+}{\sqrt{4\pi p^+}}
   \left\{ a(p^+)e^{-ip^+x^-/2} + a^\dagger(p^+)e^{ip^+x^-/2}\right\},
\ee
with the modes quantized such that
\be
[a(p^+),a^\dagger(\pp)]=\delta(p^+-\pp).
\ee
The normal-ordered light-front Hamiltonian $\Pminus=\Pfree+\Pint$ has the terms
\bea \label{eq:Pfree}
\Pfree&=&\int dp^+ \frac{\mu^2}{p^+} a^\dagger(p^+)a(p^+) \\
 && +\frac{\mu^2}{2}\int \frac{dp_1^+ dp_2^+}{\sqrt{p_1^+ p_2^+}}
    \delta(p_1^++p_2^+)\left[a^\dagger(p_1^+)a^\dagger(p_2^+)+a(p_1^+)a(p_2^+)\right],
    \nonumber
\eea
and
\bea
\Pint&=&\frac{\lambda}{2}\int \frac{dp^+d\pp}
                              {\sqrt{4\pi p^+ \pp(p^+-\pp)}}
   \left[a^\dagger(p^+)a(\pp)a(p^+-\pp) \right.\\
   && \rule{2in}{0mm} \left.
       +a^\dagger(\pp)a^\dagger(p^+-\pp)a(p^+)\right]  \nonumber \\
 && +\frac{\lambda}{6}\int\frac{dp_1^+ dp_2^+ dp_3^+}{\sqrt{4\pi p_1^+ p_2^+ p_3^+}}
     \delta(p_1^+ + p_2^+ + p_3^+) \nonumber \\
  && \rule{1in}{0mm} \times   \left[ a^\dagger(p_1^+) a^\dagger(p_2^+) a^\dagger(p_3^+)
            +a(p_1^+) a(p_2^+) a(p_3^+)\right], \nonumber
\eea
with zero-mode terms included.

The simplest approximation for $T$ is a single zero-mode creation
\be  \label{eq:T0}
T=\int_0^\infty dp^+ \sqrt{4\pi p^+}g(p^+)a^\dagger(p^+),
\ee
with $g(p^+)$ having support only at $p^+=0$ in an appropriate limit.
Momentum conservation is restored at the end of the calculation by
taking this limit.  The valence state is the bare vacuum.  The
projection $1-P_v$ is truncated to include only states with one zero mode.
The field is shifted by a constant
\be \label{eq:shift}
e^{-T}\phi e^T=\phi+[\phi,T]=\phi +\int dp^+ g(p^+) e^{-ip^+x^-/2},
\ee
in the limit that $g(p^+)\propto\delta(p^+)$.
%

\begin{figure}
\begin{center}
\begin{tabular}{c}
\includegraphics[width=11cm]{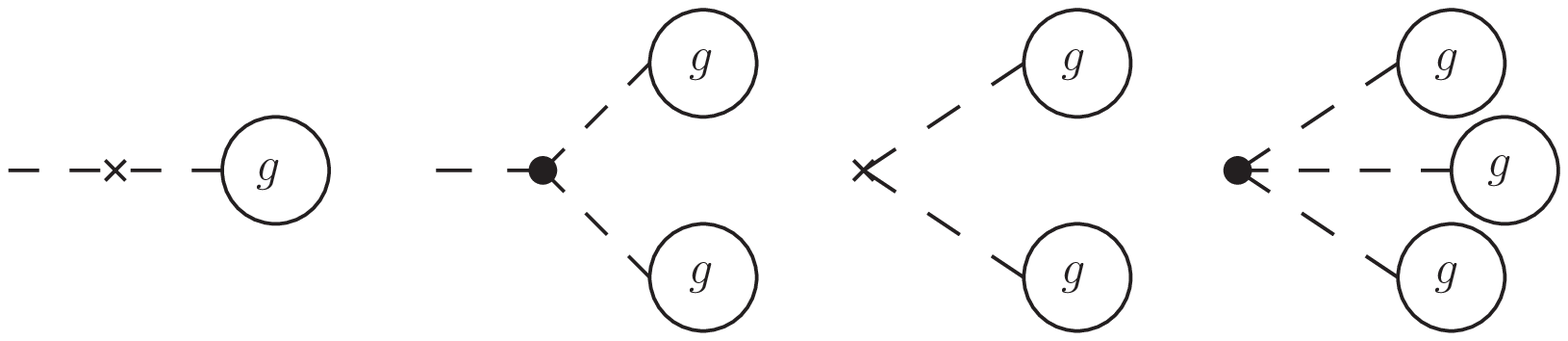} \\ (a) \\
\includegraphics[width=12cm]{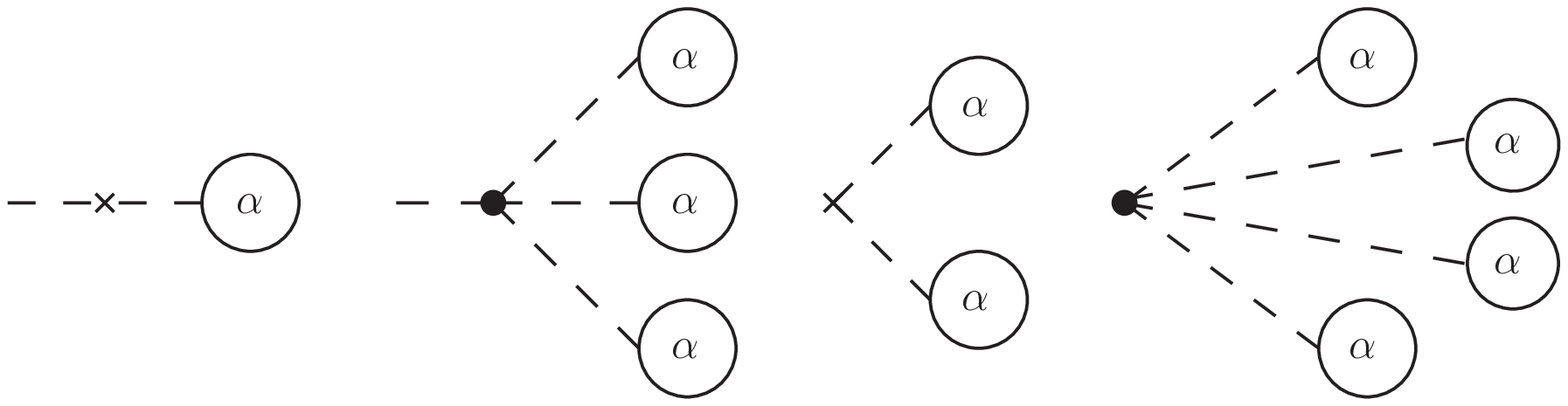} \\ (b)
\end{tabular}
\end{center}
\caption{\label{fig:EffP} Graphical representations of effective Hamiltonians for 
(a) $\phi^3$ theory and (b) $\phi^4$ theory.  The crosses represent kinetic
energy terms, and the dots, interaction terms.  Circles with $g$ and $\alpha$
represent zero-mode creation.}
\end{figure}

The effective Hamiltonian, as computed from the Baker--Hausdorff expansion, is
\bea  \label{eq:effP-}
\ob{\Pminus}&=& \sqrt{4\pi}\mu^2\int dp^+ \frac{g(p^+)}{\sqrt{p^+}}a^\dagger(p^+) \\
  && +\frac{1}{2!}4\pi\mu^2\int dp_1^+ dp_2^+ \delta(p_1^++p_2^+)g(p_1^+)g(p_2^+) 
  \nonumber \\
  && +\frac{1}{2!}\sqrt{4\pi}\lambda\int \frac{dp^+ d\pp}{\sqrt{p^+}}
       g(\pp) g(p^+-\pp) a^\dagger(p^+) \nonumber \\
  && +\frac{1}{3!}4\pi\lambda \int dp_1^+ dp_2^+ dp_3^+
     \delta(p_1^+ + p_2^+ + p_3^+) g(p_1^+)g(p_2^+)g(p_3^+), \nonumber
\eea
keeping only terms that do not annihilate the vacuum and create at most
one zero mode.  A graphical representation is given in Fig.~\ref{fig:EffP}(a).

The eigenvalue problem in the valence sector,
$P_v\ob{\Pminus}|0\rangle=P^-|0\rangle$, becomes
\bea \label{eq:phi3valence}
\lefteqn{\left[ \frac12 4\pi\mu^2\int dp_1^+ dp_2^+ \delta(p_1^++p_2^+)g(p_1^+)g(p_2^+)
                 \right.}&& \\
&&  \left.+ \frac16 4\pi\lambda \int dp_1^+ dp_2^+ dp_3^+
     \delta(p_1^+ + p_2^+ + p_3^+) g(p_1^+)g(p_2^+)g(p_3^+) \right]|0\rangle
      =P^-|0\rangle.  \nonumber
\eea
For a function $g(p^+)=\alpha\delta(p^+)$, the eigenvalue $P^-$ is
\be
P^-=\frac12 \mu^2 \alpha^2 4\pi \delta(0) +\frac16 \lambda \alpha^3 4 \pi \delta(0),
\ee
proportional to the volume
\be
\int dx^- =\lim_{p^+\rightarrow0} \int dx^- e^{ip^+x^-/2}=4\pi\delta(0).
\ee
We can then write $P^-={\cal E}^-\int dx^-$ in terms of an energy density 
${\cal E}^-=\frac12 \mu^2 \alpha^2 +\frac16 \lambda \alpha^3$.
Thus, the spectrum is unbounded from below as $\alpha$ goes
to negative infinity.

The function $g$ is determined by the auxiliary equation
$(1-P_v)\ob{\Pminus}|\phi(\ub{P})\rangle=0$,
truncated to only one zero mode,
\be \label{eq:phi3aux}
\sqrt{4\pi}\mu^2 \frac{g(p^+)}{\sqrt{p^+}}
+\frac12\sqrt{4\pi}\lambda\int_0^{p^+} \frac{d\pp}{\sqrt{p^+}}
       g(\pp) g(p^+-\pp) =0.
\ee
If this is multiplied by $\sqrt{p^+}$, a Laplace transform
$G(s)\equiv \int_0^\infty e^{-sp^+}g(p^+)dp^+$
yields
\be
\mu^2G(s)+\frac12\lambda G(s)^2=0.
\ee
The solutions are $G(s)$=0 and $-2\mu^2/\lambda$.  The 
inverse transform is then proportional to a delta function,
and we obtain the expected $g(p^+)=\alpha\delta(p^+)$ with
$\alpha=0$ and $\alpha=-2\mu^2/\lambda$.
These values of $\alpha$ correspond to the local extrema of ${\cal E}^-$;
only the global extrema at $\pm\infty$ are missed by the auxiliary equation.

With the $T$ operator truncated to one zero mode,
$|\alpha\rangle\equiv\sqrt{Z_\alpha}e^T|0\rangle$ is a coherent state~\cite{coherentstates}.
We can then minimize the vacuum energy density
$\langle\alpha|{\cal H}|\alpha\rangle$ with respect
to $\alpha$, given the Hamiltonian density
\be
{\cal H}=\frac12 \mu^2 \phi^2+\frac{\lambda}{3!}\phi^3.
\ee
The $T$ operator can be written
\be
T=\alpha\int_0^\infty dp^+ \sqrt{4\pi p^+}\Delta(p^+)a^\dagger(p^+),
\ee
where $g(p^+)$ has been replaced by $\alpha\Delta(p^+)$.  
When a specific form is needed, we take 
\be
\Delta(p^+)=\frac{1}{\epsilon}e^{-p^+/\epsilon},
\ee
defined so that $\lim_{\epsilon\rightarrow0}\Delta(p^+)=\delta(p^+)$
for integrals from 0 to $\infty$.

From the commutators
\bea
{[}T^\dagger,T]&=&4\pi\alpha^2\int dp^+ p^+ \Delta^2(p^+)\rightarrow \pi\alpha^2, \\
{[}\phi,T]&=&\alpha\int dp^+ \Delta(p^+) e^{-ip^+x^-/2}\rightarrow \alpha, \\
{[}\phi,T^\dagger]&=&\alpha\int dp^+ \Delta(p^+) e^{+ip^+x^-/2}\rightarrow \alpha,
\eea
we have, for real $\alpha$, the normalization 
$\sqrt{Z_\alpha}=e^{-\pi\alpha^2/2}$, as well as
$\phi|\alpha\rangle =\alpha|\alpha\rangle$,
$\langle\alpha|\phi=\langle\alpha|\alpha$, and
\be
\langle:\!{\cal H}\!:\rangle=\frac12\mu^2\alpha^2+\frac16\lambda\alpha^3={\cal E}^-.
\ee
The local extrema are at $\alpha=0$ and $\alpha=-2\mu^2/\lambda$,
and the global extrema at $\pm\infty$, as in the LFCC analysis.
The vacuum expectation value for the field is just 
$\langle\alpha|\phi|\alpha\rangle=\alpha$.

\section{$\phi^4$ theory}

The Lagrangian is
\be
{\cal L}=\frac12(\partial_\mu\phi)^2-\frac12\mu^2\phi^2-\frac{\lambda}{4!}\phi^4.
\ee
We again split the light-front Hamiltonian $\Pminus$ into two parts,
with $\Pfree$ as before and
\bea
\Pint&=&\frac{\lambda}{6}\int \frac{dp_1^+dp_2^+dp_3^+}
                              {4\pi \sqrt{p_1^+p_2^+p_3^+(p_1^++p_2^++p_3^+)}} \\
  &&\rule{1in}{0mm} 
    \times \left[a^\dagger(p_1^++p_2^++p_3^+)a(p_1^+)a(p_2^+)a(p_3^+)\right. \nonumber \\
  && \rule{1.25in}{0mm} \left. 
    +a^\dagger(p_1^+)a^\dagger(p_2^+)a^\dagger(p_3^+)a(p_1^++p_2^++p_3^+)\right]  \nonumber \\
 && +\frac{\lambda}{4}\int\frac{dp_1^+ dp_2^+}{4\pi\sqrt{p_1^+p_2^+}}
       \int\frac{dp_1^{\prime +}dp_2^{\prime +}}{\sqrt{p_1^{\prime +} p_2^{\prime +}}} 
       \delta(p_1^+ + p_2^+-p_1^{\prime +}-p_2^{\prime +}) \nonumber \\
 && \rule{2in}{0mm} \times a^\dagger(p_1^+) a^\dagger(p_2^+) a(p_1^{\prime +}) a(p_2^{\prime +}) 
   \nonumber \\
 && +\frac{\lambda}{24}\int\frac{dp_1^+ dp_2^+ dp_3^+ dp_4^+}{4\pi \sqrt{p_1^+ p_2^+ p_3^+p_4^+}}
     \delta(p_1^+ + p_2^+ + p_3^++p_4^+) \nonumber \\
  && \rule{0.2in}{0mm} \times   
  \left[ a^\dagger(p_1^+) a^\dagger(p_2^+) a^\dagger(p_3^+)a^\dagger(p_4^+)
            +a(p_1^+) a(p_2^+) a(p_3^+) a(p_4^+)\right]. \nonumber
\eea
The $T$ operator is approximated by 
\be
T=\alpha\int_0^\infty dp^+ \sqrt{4\pi p^+}\Delta(p^+)a^\dagger(p^+),
\ee
with $\Delta(p^+)\rightarrow\delta(p^+)$.

The effective Hamiltonian in the vacuum sector is 
\bea
\ob{\Pminus}&=& \sqrt{4\pi}\left[\mu^2\alpha+\frac16\lambda\alpha^3\right]
        \int\frac{dp^+}{\sqrt{p^+}}\Delta(p^+) a^\dagger(p^+) \\
  && +4\pi\left[\frac12\mu^2\alpha^2 +\frac{1}{24}\lambda\alpha^4\right]\delta(0). \nonumber
\eea
A graphical representation is shown in Fig.~\ref{fig:EffP}(b).

For a vacuum valence state, the valence eigenvalue problem is
\be
P_v\ob{\Pminus}|0\rangle={\cal E}^-\int dx^-|0\rangle,
\ee
with ${\cal E}^-=\frac12\mu^2\alpha^2 +\frac{1}{24}\lambda\alpha^4$.
The auxiliary equation, projected onto the
one-zero-mode sector, yields
\be
\mu^2\alpha+\frac16\lambda\alpha^3=0.
\ee
The solutions are $\alpha=0$ or $\alpha^2=-6\mu^2/\lambda$,
with $\alpha$ the vacuum expectation value for the field.
A coherent-state analysis yields the same results~\cite{LFCCzeromodes}.

For the wrong-sign case, where $\mu^2\rightarrow-\mu^2$,
the solution $\alpha=\pm\sqrt{6\lambda}/\mu$ corresponds to a shift
of the field $\phi$.  This shift brings the Hamiltonian density to a minimum
and shows that the inclusion of a zero mode in the LFCC $T$ operator is
an important step.

In general, the effective Hamiltonian will have terms that
mix Fock states with odd and even numbers of particles, which is
characteristic of broken symmetry.
For example, a commutator that contributes to the Baker--Hausdorff
expansion of $\ob{\Pminus}$ is
\bea
\label{eq:PintT}
{[}\Pint,T]&=& \frac{\lambda\alpha}{2}\int 
   \frac{dp_1^+ dp_2^+ dp^+}{\sqrt{4\pi p_1^+ p_2^+ p^+}}
     \delta(p^+-p_1^+ -p_2^+) \\
 && \rule{0.5in}{0mm} \times
     \left[a^\dagger(p^+)a(p_1^+) a(p_2^+)
                   +a^\dagger(p_1^+) a^\dagger(p_2^+) a(p^+)\right],  \nonumber
\eea
which changes particle number by one.

\section{Summary}  \label{sec:summary}

In light-front quantization, the vacuum is
trivial without zero modes.
The LFCC method, which solves the light-front
Hamiltonian eigenvalue problem nonperturbatively, can be extended to
include zero modes.
In $\phi^3$ and $\phi^4$ theories, we have shown that
this provides for the standard vacuum expectation value and is
consistent with a variational coherent-state analysis.
A four-zero-mode calculation in $\phi^4$
theory is underway, to compute the critical coupling for dynamical 
symmetry breaking.
Another accessible application is a
nonperturbative calculation of the Higgs mechanism.

\acknowledgments
This work was done in collaboration with J.R. Hiller
and supported in part by the US Department of Energy and the Minnesota Supercomputing Institute.

\end{document}